\documentclass[%
 reprint,
 amsmath,amssymb,
 aps,
prb,
]{revtex4-2}

\usepackage{graphicx}
\usepackage{dcolumn}
\usepackage{bm}
\usepackage{color}
\usepackage{comment}

\begin{document}

\preprint{APS/123-QED}

\title{Ultra-lightweight multi-functional gyroid-based metamaterial for simultaneous sound insulation, vibration suppression, and impact mitigation in all directions}

\author{Evan Kluge}
\author{Osama R. Bilal}%
 \email{osama.bilal@uconn.edu}
 \affiliation{School of Mechanical, Aerospace, and Manufacturing Engineering, University of Connecticut, Storrs, USA}

\date{\today}

\begin{abstract}

Multi-functional materials are increasingly essential in many applications due to their ability to serve multiple purposes, simultaneously. Metamaterials can offer tailored functionality based on carefully designed tessellating basic building blocks, or unit cells, in space. These unit cells are most commonly designed for a single objective, for example, to mitigate an impact, insulate sound, or dampen vibrations. However, a metamaterial that can mitigate impact, attenuate elastic vibrations, insulate airborne sound, while being ultra-light and ultra-stiff remains elusive. Here, we present three gyroid-based metamaterial designs, which are numerically and experimentally examined, that can simultaneously serve all these functions in all directions. Our designs have band gaps reaching 60\% with as low as 367 Hz starting band gap frequency, while being up to 77\% lighter than a homogeneous block of the same material. Our findings highlight the potential of metamaterials for applications that require wave manipulation, mechanical resilience, and impact mitigation with limited mass and volume, simultaneously.

\end{abstract}

\maketitle


\section{\label{sec:level1} Introduction}

Many applications require multifunctional materials that are lightweight, stiff, capable of mitigating impact, damping vibrations, and blocking noise, all within a limited volume. However, effectively achieving all these properties, at the same time, remains a significant challenge, mainly due to conflicting design objectives. Metamaterials offer a unique opportunity to tailor the mechanics of materials based on their geometry rather than their chemical composition. Numerous studies in the literature tackle the design of metamaterials' basic building blocks, or unit cells, for a single objective function. For example, some metamaterials have been proposed for impact mitigation \cite{almahri2021evaluation, smith2024tunable, gartner2024geometric, andrew2021energy, supian2018hybrid}. Other metamaterials were designed to attenuate airborne noise either through scattering \cite{aravantinos2023complete, shendy2024extensive, aravantinos2022phononic, konstantopoulou2019wide} or resonance \cite{bicer2021broad, domingo2021acoustic, li2021microlattice}. Different unit cells were designed to control elastic vibrations through solids, either by means of scattering \cite{bilal2011optimization,an2019three, gerard2021three, zhang2023vibration, fei2020three}, resonance \cite{liu2000locally, pham2024composite, alhammadi2021numerical, yu2023integrated} or a combination of both \cite{elmadih2019multidimensional, bilal2021experimental,hou2024hybrid, jiang2022multifunctional, matlack2016composite, roshdy2023tunable,li2022mitigation}.

\begin{figure}[b]
\includegraphics{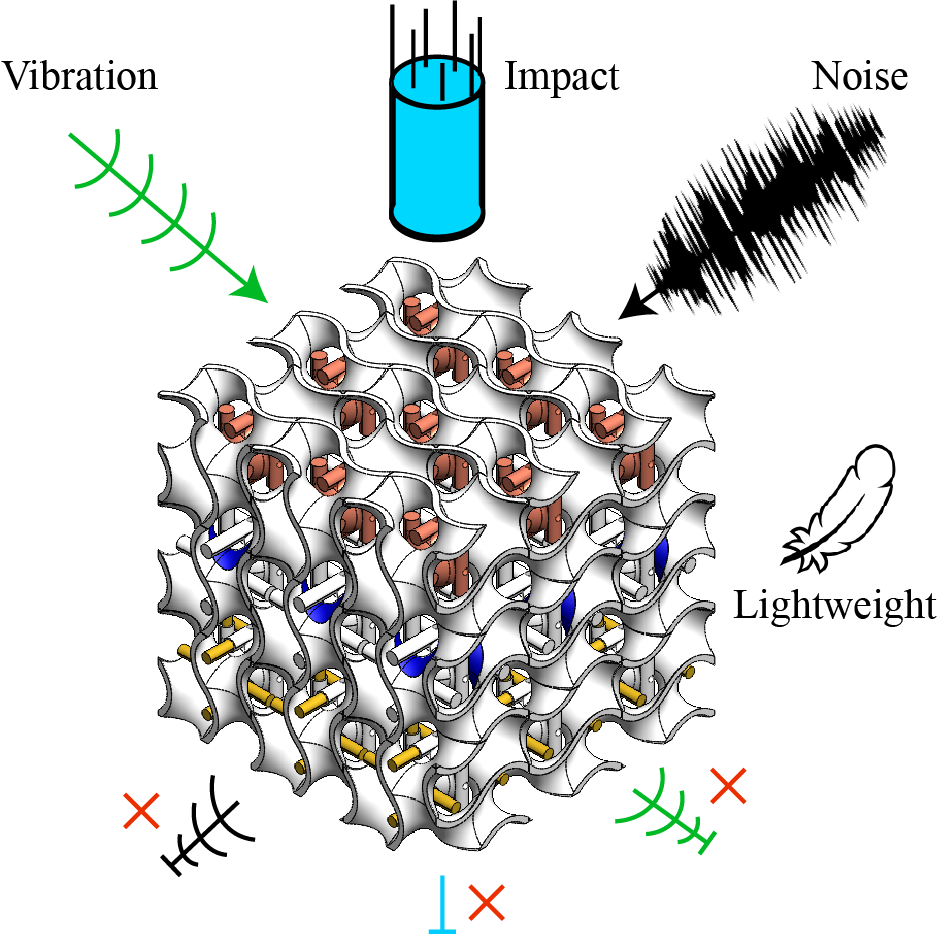}%
\caption{\label{fig:concept}\textbf{ Metamaterial concept.} Schematic of an ultra-lightweight metamaterial used for simultaneous sound and vibration attenuation and impact mitigation.}
\end{figure}

An emerging design approach \cite{bilal2018architected} proposes the combination of elastic and acoustic metamaterials into a single, vibro-acoustic unit cell  with varying levels of periodicity \cite{ kheybari2022tunable, aravantinos2021acoustoelastic, jiang2019lightweight, li2022local,elmadih2021metamaterials, gu2023lightweight, wang2024elastoacoustic, zhang2024multifunctional, xu2024metamaterial,fu2025simultaneous,zeng2024nonlocal}. In addition, a few studies introduced multi-functionality through the combination of elastic and acoustic wave attenuation, along with impact mitigation \cite{li2021microlattice, zhang2023vibration, li2022mitigation, jiang2022multifunctional}.   Triply periodic minimal surfaces (TPMS) have gained significant attention for their unique geometric and mechanical properties. Metamaterials based on TPMS offer a unique pathway for integrating sound insulation, vibration attenuation, and impact mitigation in a single structure. The gyroid, a TPMS discovered by Alan H. Schoen in 1970 \cite{schoen1970infinite}, is characterized as an intersection-free infinite periodic surface with zero-mean curvature, minimizing the local area of any small patch defined within its boundaries \cite{maskery2017compressive, abueidda2019mechanical, ramos2022response}. Schoen described the gyroid as lacking straight or planar lines of curvature, with its symmetry group notably devoid of mirror reflections and its rotational axes not lying within the surface itself. These geometric qualities lend the gyroid its inherent high stiffness-to-weight ratio and isotropic behavior, making it an ideal candidate as the basis for a multifunctional metamaterial.

Several studies have numerically determined the mechanical characteristics of gyroids, often alongside other TPMS structures \cite{al2018microarchitected, al2019multifunctional, maskery2018insights}. Experimental investigations, including quasi-static compression tests, have further validated the mechanical behavior of gyroids and other TPMS designs, highlighting their potential for structural applications \cite{qiu2024experimental, nazir2024design, li2024enhancing}. Experimental comparisons have revealed differences in the mechanical performance of sheet-based and strut-based gyroid lattices, with sheet structures exhibiting isotropic behavior across density ranges and enhanced energy absorption \cite{li2019comparison}.  Additional work focused on the manufacturability and mechanical testing of 3D-printed gyroids and other TPMS designs, emphasizing their suitability for multifunctional applications \cite{al2019multifunctional}. Recent advances in TPMS design tools have expanded their applicability. The MSLattice software, for example, allows precise manipulation of gyroid parameters such as thickness, periodicity, and volume fraction, enabling customization of structures for specific applications \cite{al2021mslattice}. The static properties of gyroid lattices such as Young's modulus and Poisson's ratio are influenced by proportion, scale, and shape, as well as the volume fraction of the unit cell \cite{jones2023investigating, al2017mechanical, al2018nature, devalk2021poisson}. Beyond static properties, the gyroid has been evaluated for dynamics-based applications, including impact mitigation and energy absorption. Experimental and numerical studies have investigated the behavior of gyroids under impact loading, demonstrating their ability to effectively dissipate energy and absorb impacts \cite{tilley20243d, alemayehu2024enhanced, rostro2024additive, vrana2016impact, ramos2022response, ramos2022response, almahri2021evaluation}. Numerical and experimental studies also highlight their structural utility as an infill for additive manufacturing \cite{bean2022numerical, stromberg2021optimal}. The gyroid further demonstrates potential for manipulating wave propagation. Numerical studies have shown its ability to exhibit three-dimensional acoustic band gaps that vary with lattice thickness or size \cite{lu2024insights, silva2024investigation, lu2021acoustic}. Similarly, one-dimensional elastic band gaps have been identified in gyroid structures, highlighting their potential for vibration attenuation in engineered systems \cite{elmadihmechanical}. In the electromagnetic domain, both the sheet and strut-based gyroids have been shown to exhibit band gaps as photonic materials \cite{michielsen2003photonic, martin1999self, peng2016three}.

A multifunctional metamaterial with the ability to mitigate impact, attenuate both elastic and acoustic waves in all directions, while being ultra-light and ultra-stiff remains elusive. In this work, we design, realize, and experimentally validate three-dimensional gyroid-based metamaterials that can mitigate impact loads, attenuate mechanical vibrations propagating through the solid, insulate from noise propagating through the air, while being ultra-light and ultra-stiff (Fig.\ref{fig:concept}).

\begin{figure}
\includegraphics{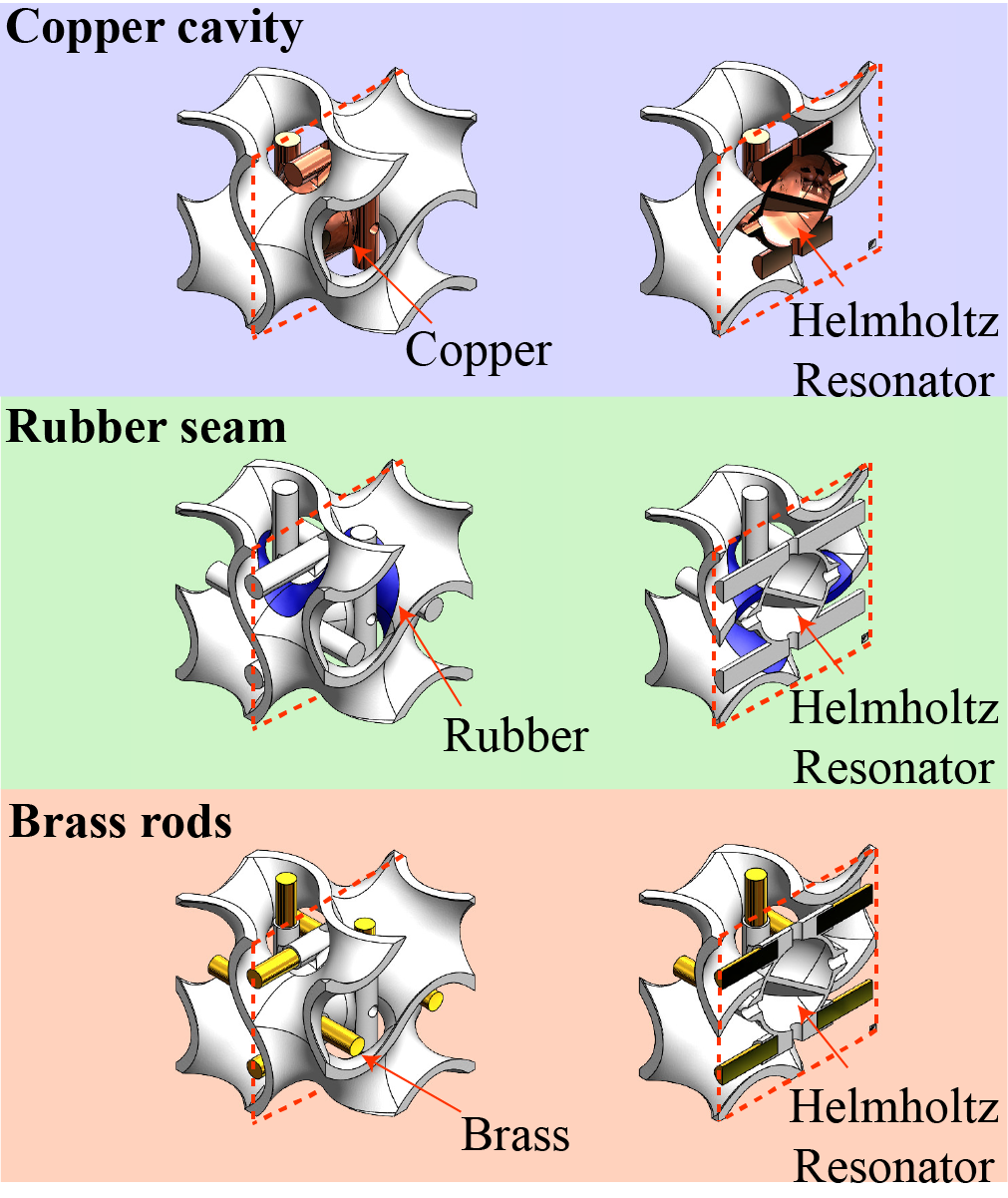}%
\caption{\label{fig:3 Unit Cells}\textbf{Three design choices.} The copper cavity design in the purple region, the rubber seam design in the green region, and the brass rods design in the orange region.}
\end{figure}

\begin{figure}[b]
\includegraphics{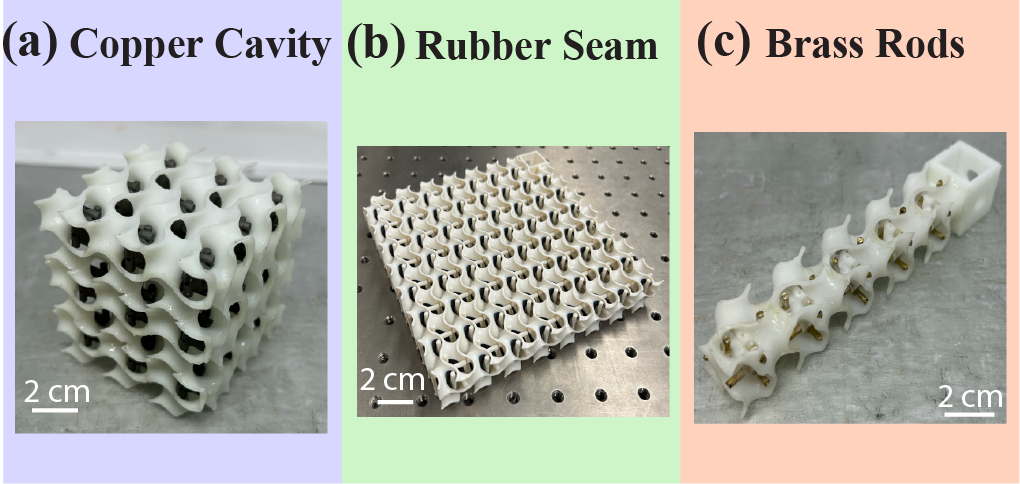}%
\caption{\label{fig:Sample Fabrication}\textbf{Fabricated designs.} (a) The copper cavity design. (b) The rubber seam design.  (c) The brass rods design.}
\end{figure}

\begin{figure*}
\includegraphics{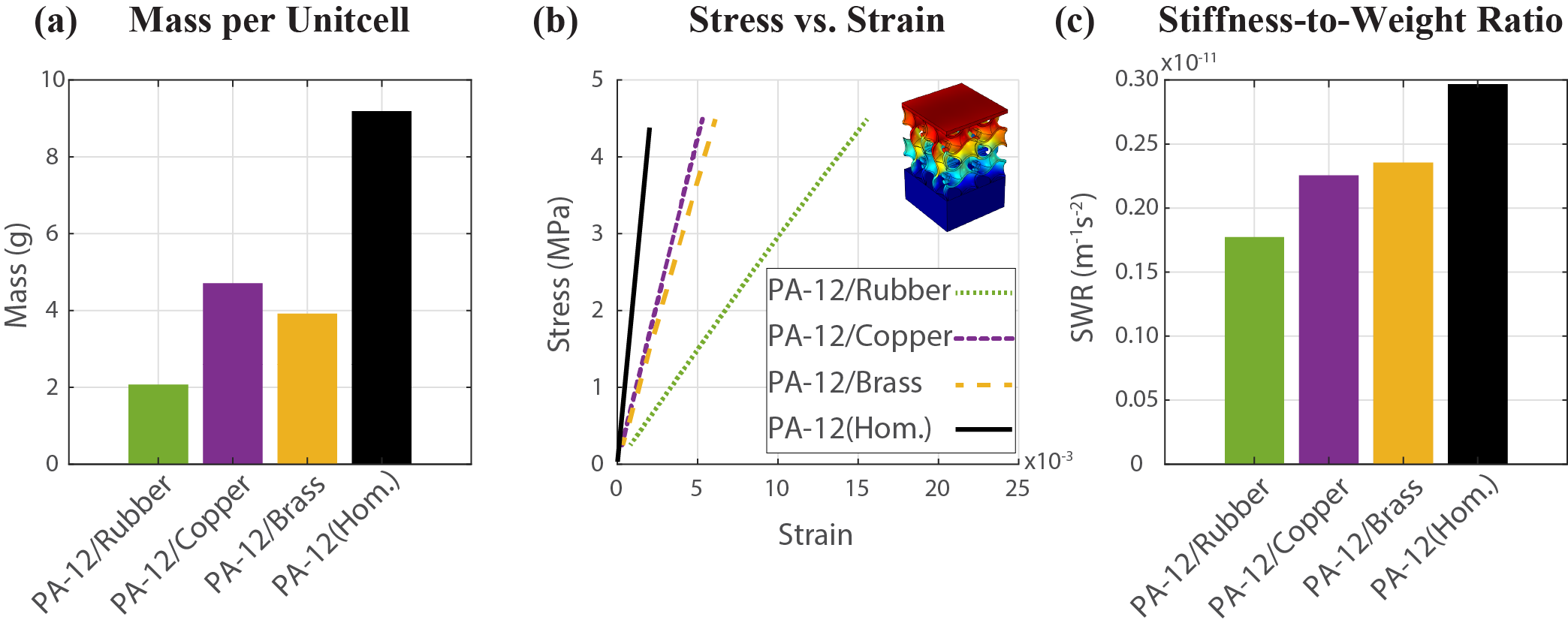}%
\caption{\label{fig:Stress and SWR}\textbf{Quasi-static analysis.} (a) The mass of each design per unit cell. (b) Stress versus strain plot for each design compared to a homogeneous block of PA-12. (c) Normalized stiffness-to-weight ratio (SWR) plot compared to a homogeneous block of PA-12.}
\end{figure*}

\section{Design Methodology}

To design our metamaterials, we start with the conventional sheet gyroid structure as it has excellent stiffness-to-weight ratio (SWR) that is comparable to a homogeneous block of the same base material. These gyroid structures can be generated by using the formula:
\begin{multline} \label{eq: 1}
    sin(\frac{2\pi}{a}X)cos(\frac{2\pi}{a}Y)+sin(\frac{2\pi}{a}Y)cos(\frac{2\pi}{a}Z)
    \\+sin(\frac{2\pi}{a}Z)cos(\frac{2\pi}{a}X)=c, 
\end{multline}
where \textit{a} is the unit cell size and \textit{c} is the volume fraction of the material within the unit cell. We utilize the same gyroid structure made out of nylon (PA-12) in all three proposed designs. To introduce a mechanism to attenuate waves within our designs, we rely on local resonating elements for both acoustic emissions and mechanical vibrations. For acoustic waves propagating through the air, we embed a Helmholtz-like resonator at the center of each of the three proposed unit cells (Fig.\ref{fig:3 Unit Cells}). All Helmholtz cavities are composed of two identical semi-spherical chambers, each with three openings. Such air cavities within the unit cells introduce local resonance for an incoming acoustic wave. We also use resonances for elastic waves, but utilize three different design approaches. In the first design, we make the solid chamber of the air cavity (i.e., the Helmholtz resonator) out of copper, a much denser material than the gyroid base material (i.e., PA-12). We also attach six short cylindrical beams to the solid chamber to enhance resonance. In the second design, we make the solid chamber out of PA-12 (i.e., the same material as the gyroid), however, we attach it to the gyroid structures through a rubber seam, to allow the entire chamber to resonate elastically. In the third and last design, both the chamber and the gyroid structures are fully made out of the same material (PA-12), with added brass rods inserted into the side beams of the chamber. Each of the three design concepts has implications on the capacity of the metamaterial for mitigating impact. For design 1, the copper chamber adds stiffness to the gyroid structure making it comparable to the stiffness of a PA-12 homogeneous block under impact while being much lighter. For design 2, the rubber seam reduces both the stiffness of the structure and its mass, making it an ideal candidate for impact mitigation. Design 3 with the brass rods is an intermediate state between design 1 and design 2.

\section{Sample Fabrication}

In order to validate our numerical predictions, we fabricate multiple prototypes of our designs. While each design provides a unique opportunity to mitigate impact, attenuate vibrations, insulate sound, and maintain high stiffness-to-weight ratio, these designs present different challenges in manufacturing. To show the versatility of our design approaches and the manufacturability of the proposed concepts, we explore different fabrication methods for each design (Fig.\ref{fig:Sample Fabrication}). For the design with the copper cavity and plastic gyroid structure, the gyroid is additively manufactured using stereolithography (Formlabs Form 3) with rigid 4000 resin. The metal core is additively manufactured by powder laser sintering. For the assembly of the final prototypes, we print the gyroid unit cell structures in halves. We place the metal core between the two halves and coat the interior surface of each half with the same resin and cure it, to create one homogeneous structure (Fig.\ref{fig:Sample Fabrication}a). We note that during the fabrication we use stainless steel 316-L instead of copper as both have comparable densities. For the design with the rubber seam, we utilize polyjet additive manufacturing technology, allowing for the printing of both the plastic and the rubber connections simultaneously. The gyroid structure and the core chamber are printed with  Vero-Ultra White and the rubber seam is printed with Agilus-Black 85A (Fig.\ref{fig:Sample Fabrication}b). For the design with the brass rods, the gyroid is additively manufactured using stereolithography (Formlabs Form 3) with rigid 4000 resin. We then insert the brass rods in the printed hollow tube in each unit cell.  Once each rod is inserted in the correct position within one unit cell, the cells are joined together with the same resin as the gyroid structure and cured to create a homogeneous array (Fig.\ref{fig:Sample Fabrication}c).

\section{Quasi-static Properties}

To characterize the quasi-static properties of our three designs, we determine the stiffness-to-weight ratio (SWR) defined as the ratio between the modulus of elasticity, $E$, and the mass, \textit{m}, of each unit cell (Fig.\ref{fig:Stress and SWR}a). We start by considering the mass of each unit cell. The mass of copper cavity unit cell is 4.71 g, for the rubber seam unit cell is 2.07 g, and for the brass rods unit cell is 3.92 g. The mass of a homogeneous PA-12 block with the same unit cell dimensions is 9.2 g. Comparing our designs against the homogeneous block, the copper cavity design is 48.8\% lighter, the rubber seam design is 77.5\% lighter, and the brass rods design is 57.5\% lighter. 

We also consider the stiffness of our designs, by calculating the slope of the stress-strain curve under static loading conditions using the finite element method (Fig.\ref{fig:Stress and SWR}b). As expected, the homogeneous block is the most stiff followed by the design with the copper cavity, the brass rods, and finally the rubber seam. The rubber seam  decreases the overall stiffness because of its compliance to deform, whereas the copper cavity increases the stiffness because of the contribution of the copper core.

Taking into account both the stiffness and the weight of a unit cell of each design, allows for a fair comparison between them (Fig.\ref{fig:Stress and SWR}c). The homogeneous block has the highest SWR of 0.297. The highest performing design among our three designs is the unit cell with the brass rods with a SWR of 0.236, followed by the copper cavity, 0.226, and the rubber seam, 0.177. As expected, the homogeneous block is the most stiff for its weight, but does not have the ability to mitigate impact, insulate vibrations, or attenuate noise. The three gyroid configurations however, have comparable SWRs, while exhibiting the ability to serve multiple purposes by mitigating impact,  attenuating vibration, and insulating sound, simultaneously, as we show in the following sections.

\section{Dynamic Properties}
In addition to the quasi-static analysis of our proposed designs, we consider their performance under dynamic loading. We investigate their utility in (1) impact mitigation, (2) mechanical vibration attenuation, and (3) sound wave insulation.

\subsection{Impact Mitigation}

\begin{figure}
\includegraphics{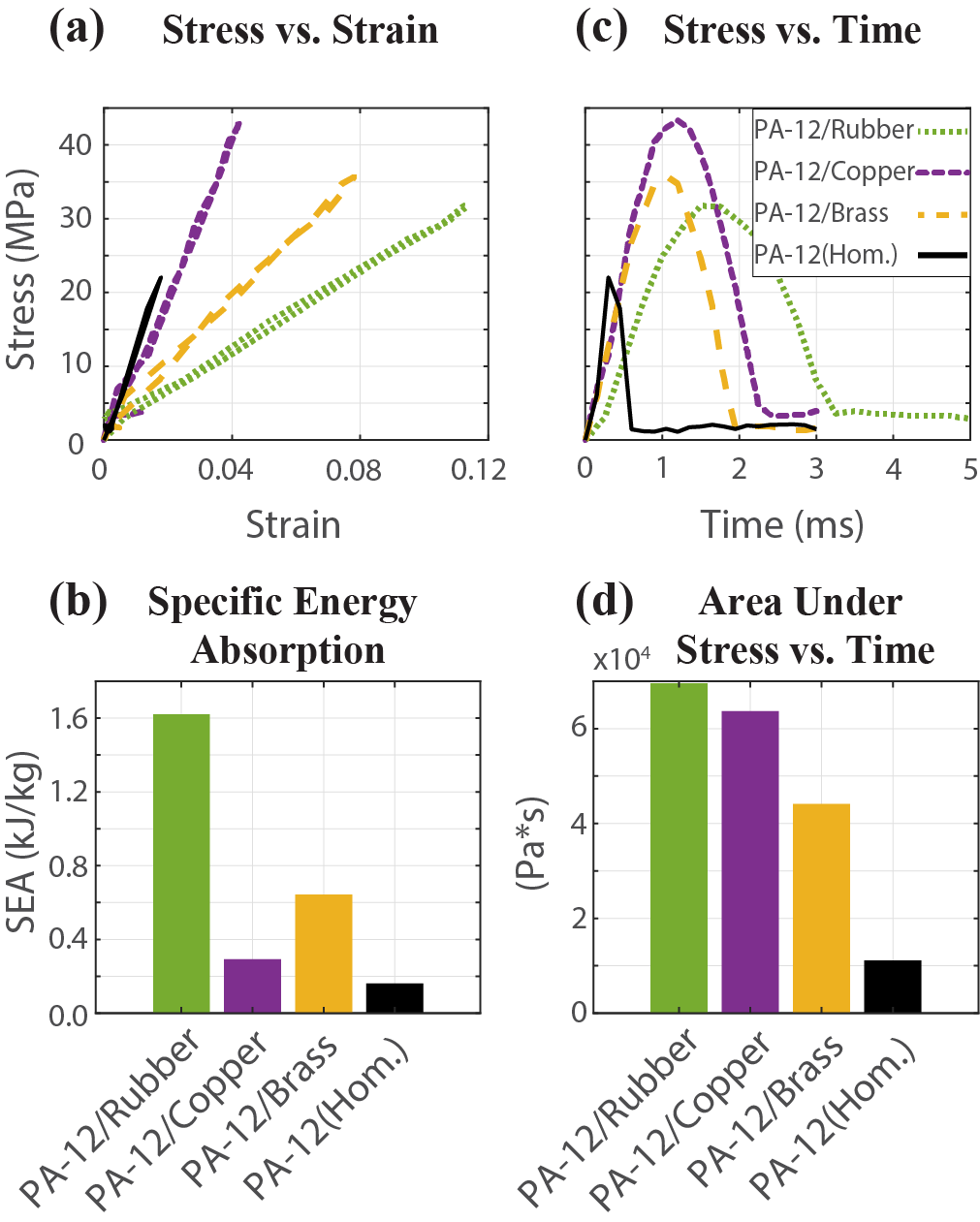}%
\caption{\label{fig:Impact Analytical}\textbf{Dynamic impact analysis.} (a) Stress versus strain plot during a 5 m/s impact. (b) Specific energy absorption. (c) Stress versus time plot during a 5 m/s impact. (d) Area under the stress versus time plot.}
\end{figure}

We start our dynamic analysis by considering the capacity of the proposed design to mitigate an  impact by adding a dynamic load at the top edge of a three-dimensional (3D) lattice composed of each design. In general, a material that is best suited for impact mitigation exhibits low peak stress amplitude under a given load and a large area under the stress-time curve, indicating a prolonged duration of impact. Therefore, we examine both the stress-strain and stress-time curves for each lattice with an impact velocity of 5 m/s. 

By inspecting the slope of the stress-strain curves for all three designs, (Fig.\ref{fig:Impact Analytical}a), the copper cavity design shows the highest stiffness, followed by the design with the brass rods, then the design with the rubber seam. As expected, the homogeneous block is the stiffest compared to all three designs. To determine the energy absorption capacity of each design, we calculate the area under the stress-strain curve. The specific energy absorption (SEA) is calculated by multiplying the area under the curve, $EA$, by the unit cell volume, $V$, all divided by the unit cell mass, $m$: \(SEA = \frac{EA*V}{m}\) (Fig.\ref{fig:Impact Analytical}b). The design with the highest capacity for energy absorption by a significant margin is the design with the rubber seam, followed by the design with the brass rods, the design with the copper cavity, and finally the homogeneous block. 

An additional metric for energy absorption is the width of the stress-time plot. For impact mitigation, materials with a lower peak stress and a wider stress-time profile are generally better (Fig.\ref{fig:Impact Analytical}c). The design with the widest stress-time profile is the design with the rubber seam, followed by the design with the copper cavity, then the design with the brass rods, and finally the homogeneous block. To better analyze the stress-time plot, we plot the area under the curves for each design. The greater the area under the curve, the better the material is in impact mitigation (Fig.\ref{fig:Impact Analytical}d). Overall, based on all impact calculations, the design with the rubber seam looks the most promising for impact mitigation. The design's ability to deform under dynamic loading, while maintaining stiffness, makes it a more suitable candidate for impact mitigation applications.

\subsection{Elastic and Acoustic Insulation}

\begin{figure}[b!]
\includegraphics{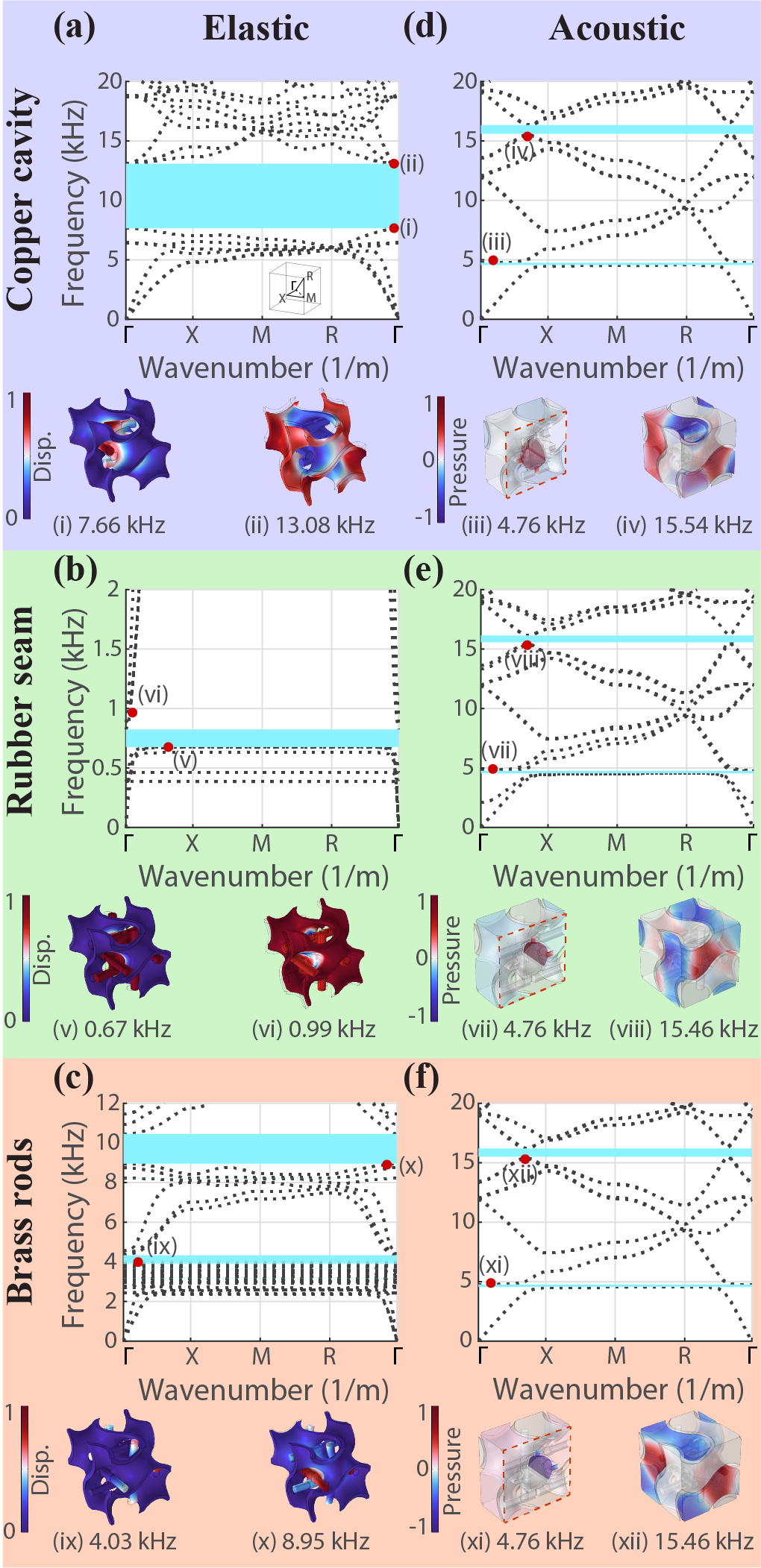}%
\caption{\label{fig:Dispersion Curves}\textbf{3D dispersion curves.} Dispersion curves in the elastic domain for (a) copper cavity design,  (b) rubber seam design, and (c) brass rods design. Dispersion curves in the acoustic domain for (d) copper cavity design,  (e) rubber seam design, and (f) brass rods design.  The teal shaded regions represent the band gap frequency ranges. Subplots (i-xi) represent select mode shapes at the edges of the band gap frequency ranges. }
\end{figure}

To validate our design principles, we first consider a single unit cell repeated infinitely in space. We numerically simulate the wave propagation characteristics of the three designs using the finite element method. We apply Bloch periodic boundary conditions in all directions utilizing COMSOL Multiphysics (v6.2) and formulate an eigenvalue problem correlating wavenumber to frequency. The solution is the wave function \(u(x,\kappa;t)=\tilde{u}(x)exp{[i(\kappa^\intercal x-\omega)]}\), where $\tilde{u}$ is the Bloch displacement vector, \textit{x} is the position vector, $\kappa$ is the wavenumber, $\omega$ is the frequency, and $t$ is time. 
We plot the resulting solutions for each unit cell design for both the solid domain for elastic waves and the surrounding air for airborne sound. All the dispersion curves are calculated for the wavenumbers spanning the high symmetry points \(\Gamma-X-M-R-\Gamma\), which accounts for the irreducible Brillouin zone, or the smallest repeatable portion in the reciprocal space (Fig.\ref{fig:Dispersion Curves}a-inset). All of the three designs include full 3D band gaps in their frequency spectrum. In other words, our metamaterial designs can attenuate elastic vibrations and airborne sound in all directions and wave polarizations (i.e., in-plane and out-of-plane). 

For the elastic wave calculations, we consider only the solid part of the three designs. For design 1 with the copper cavity, the hybridization between Bragg-scattering and resonance opens the largest band gap of all the three design configurations between 7.66 and 13.08 kHz (Fig.\ref{fig:Dispersion Curves}a). For design 2, with the rubber seam, the pure local resonance opens the lowest band gap of all three designs between 0.67 and 0.82 kHz (Fig.\ref{fig:Dispersion Curves}b). The band gap is a direct result of the rubber seam connecting the gyroid structure to the resonating solid chamber. For design 3 with the inserted brass rods, there exists two band gaps as a result of resonance for the band gap ranging from 4.03 to 4.33 kHz and additional band gap between 8.96 and 10.45 kHz (Fig.\ref{fig:Dispersion Curves}c). 

For the acoustic wave calculations, we numerically model the negative of the solid part of the three designs to be made out of air. All three designs have the same air chamber design and therefore, share the same lower band gap frequency range between 4.56 and 4.76 kHz (Fig.\ref{fig:Dispersion Curves}d,e,f). This band gap is independent of the gyroid structure and the material of the chamber. The upper band gaps from 15.5 and 16.2 kHz also show great similarity, varying only by $\sim$ 100 Hz. The higher frequency band gaps are due to Bragg-scattering, which is affected by the variation between the three designs. 

\subsection{Parametric Optimization}

\begin{figure}
\includegraphics{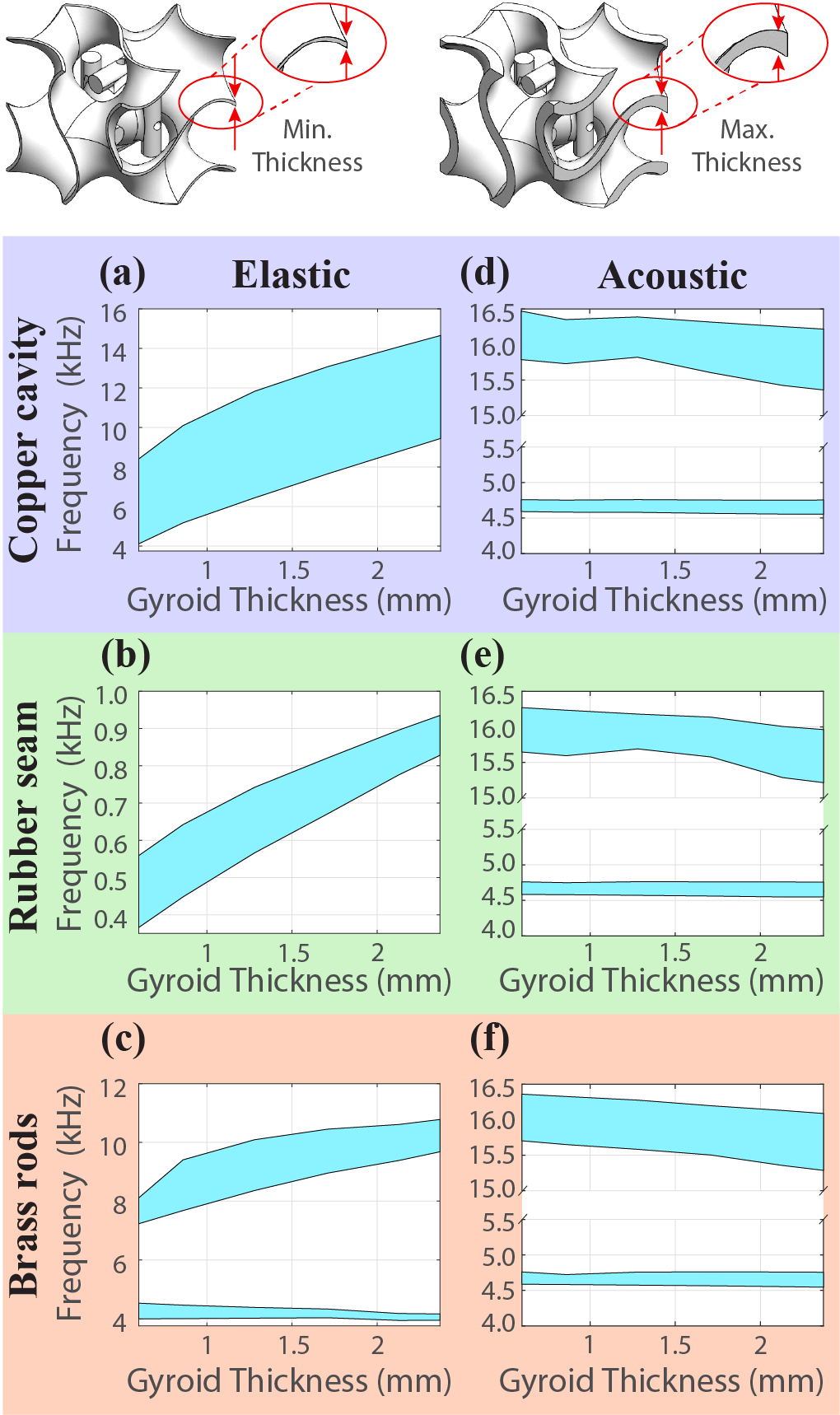}%
\caption{\label{fig:Thickness Comparisons}\textbf{Gyroid thickness parametric comparisons.} (a) Copper cavity design band gap region in the elastic domain. (b) Rubber seam design band gap region in the elastic domain. (c) Brass rods design band gap region in the elastic domain. (d) Copper cavity design band gap region in the acoustic domain. (e) Rubber seam design band gap region in the acoustic domain. (f) Brass rods design band gap region in the acoustic domain.}
\end{figure}

The design space for the gyroid based metamaterials is rich with various opportunities for different functionalities and operational frequencies. For example, the effective parameters of our three designs can be varied by changing (1) the thickness of the gyroid structure, (2) the diameter of the opening to the air chamber, and (3) the diameter of the air chamber. We systematically vary each of these three design parameters to explore their effect on the position and width of the band gap frequencies for both elastic and acoustic waves. For elastic wave attenuation, changing the gyroid structure thickness between 0.6 and 2.37 mm has the greatest influence on the elastic band gap frequency range (Fig.\ref{fig:Thickness Comparisons}). For all three designs, as the gyroid thickness increases, the elastic band gap central frequency increases due to the increase in stiffness. For the design with the copper cavity, the band gap central frequency region increases by $\sim$ 6 kHz with the lower edge of the band gap increasing from $\sim$ 4 to 10 kHz, and the upper edge of the band gap increasing from $\sim$ 8 to 14 kHz (Fig.\ref{fig:Thickness Comparisons}a). For the design with the rubber seam, the central frequency of the band gap increases by $\sim$ 0.4 kHz with the lower edge of the band gap increasing from $\sim$ 0.4 to 0.8 kHz and the upper edge of the band gap increasing from $\sim$ 0.55 to 0.95 kHz (Fig.\ref{fig:Thickness Comparisons}b). For the design with the brass rods, there are two band gaps showing different trends as we increase the gyroid structure thickness. The central frequency of the higher band gap increases by $\sim$ 2 kHz with the upper edge increasing from $\sim$ 7.5 to 9.5 kHz and the upper edge increasing from $\sim$ 8 to 10.5 kHz (Fig.\ref{fig:Thickness Comparisons}c). The lower band gap, however, does not change significantly as we increase the thickness of the gyroid structure mainly due to the independence of the local resonance from the thickness of the gyroid structure in this configuration.

For acoustic wave attenuation, there are two band gaps within all three designs' frequency spectrum (Fig.\ref{fig:Thickness Comparisons}d,e,f). The lower gap stems from resonance, while the higher one is caused by scattering. The change in the thickness of the gyroid structure has no effect on the lowest band gap frequency range. However, the central frequency of the upper band gap for all three designs is slightly decreased by the increase of the gyroid structure thickness.

\begin{figure}
\includegraphics{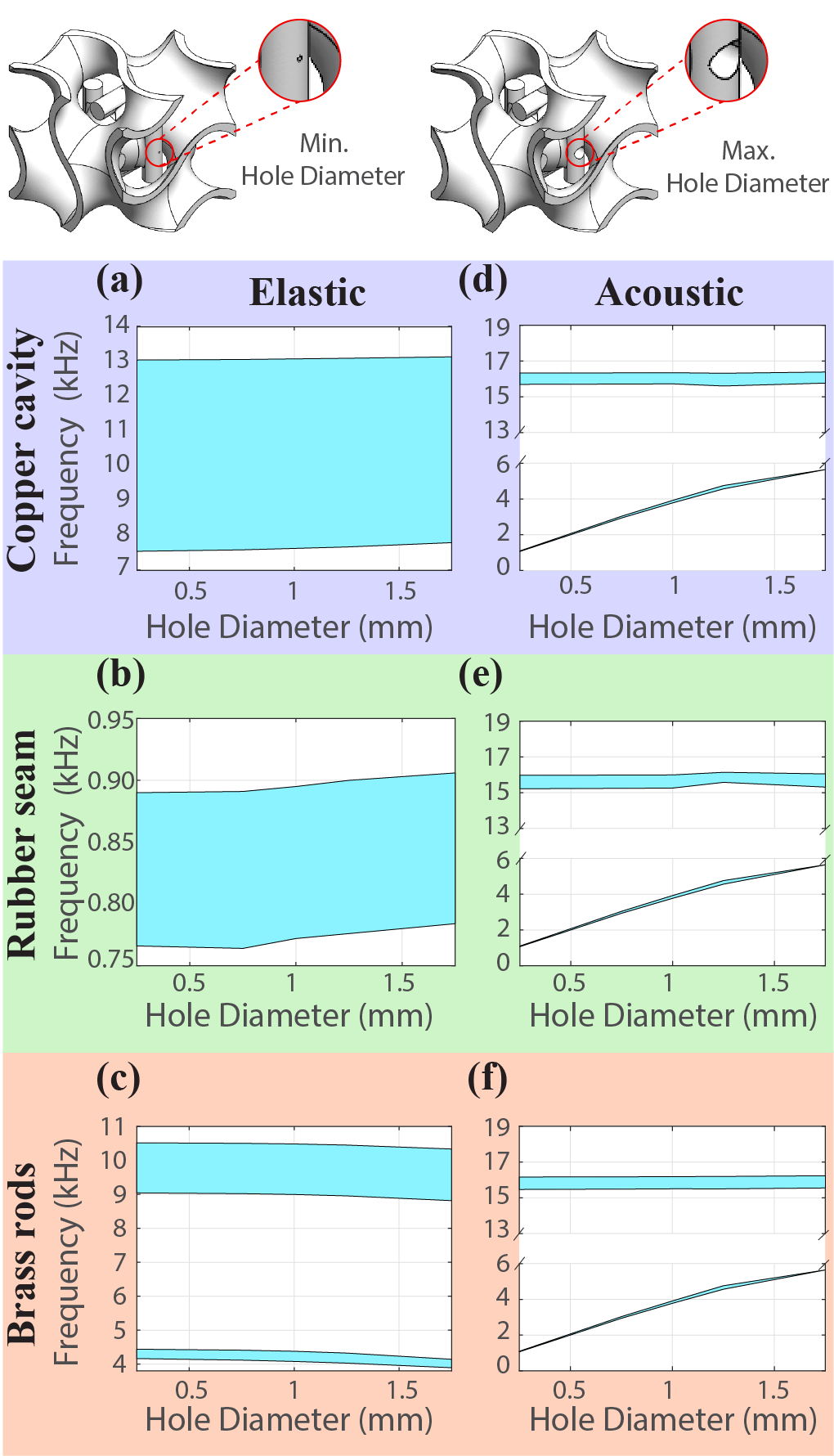}%
\caption{\label{fig:Hole Diameter Comparisons}\textbf{Helmholtz interior hole diameter parametric comparisons.} (a) Copper cavity design band gap region in the elastic domain. (b) Rubber seam design band gap region in the elastic domain. (c) Brass rods design band gap region in the elastic domain. (d) Copper cavity design band gap region in the acoustic domain. (e) Rubber seam design band gap region in the acoustic domain. (f) Brass rods design band gap region in the acoustic domain.}
\end{figure}

In addition to exploring the influence of the gyroid thickness on the designs' dynamic characteristics (with predominant effect on the elastic band gap frequencies), we also explore the change in diameter size for the air chamber opening (Fig.\ref{fig:Hole Diameter Comparisons}). As expected, the change in the hole diameter has no to little effect on the elastic wave attenuation for all three designs. For the design with the copper cavity and the rubber seam, the band gap central frequency increases by $\sim$ 100 Hz due to the slight decrease in the mass of the resonating rods (Fig.\ref{fig:Hole Diameter Comparisons}a,b). For the design with the brass rods, the central frequency of both band gaps decreases by $\sim$ 250 Hz (Fig.\ref{fig:Hole Diameter Comparisons}c). As the hole diameter increases, the stiffness of the base of the resonating rods decreases, leading to the reduction of the band gap central frequency.

For acoustic wave attenuation, when varying the diameter of the opening of the air chamber (i.e., Helmholtz-like cavity), we observe the most impact on the resonance-based acoustic band gaps in all three designs (Fig.\ref{fig:Hole Diameter Comparisons}d,e,f). The cavity at the center of each unit cell has identical parameters in the sense of the Helmholtz resonance equation, albeit with different solid materials:
\[
f = \frac{c}{2\pi} \sqrt{\frac{A}{V \cdot L_\text{eff}}},
\] where $c$ is the speed of sound in air, $A$ is the cross sectional area of the Helmholtz opening using the opening diameter $d$ (\(A = \frac{\pi d^2}{4} \)), $V$ is the volume of the interior section, and \(L_\text{eff} \) is the length of the neck multiplied by an end correction term \cite{selamet1994theoretical}. This translates to all three designs having the same lower band gap frequency region stemming from resonance. As the hole diameter increases, the band gap central frequency increases significantly by $\sim$ 5 kHz from $\sim$ 1 to 6 kHz. When the opening diameter is less than 0.25 mm or greater than 1.75 mm, the band gap closes. The frequency of the second band gap is unaffected by the change in the opening diameter in all three designs since this gap stems from scattering.

\begin{figure}
\includegraphics{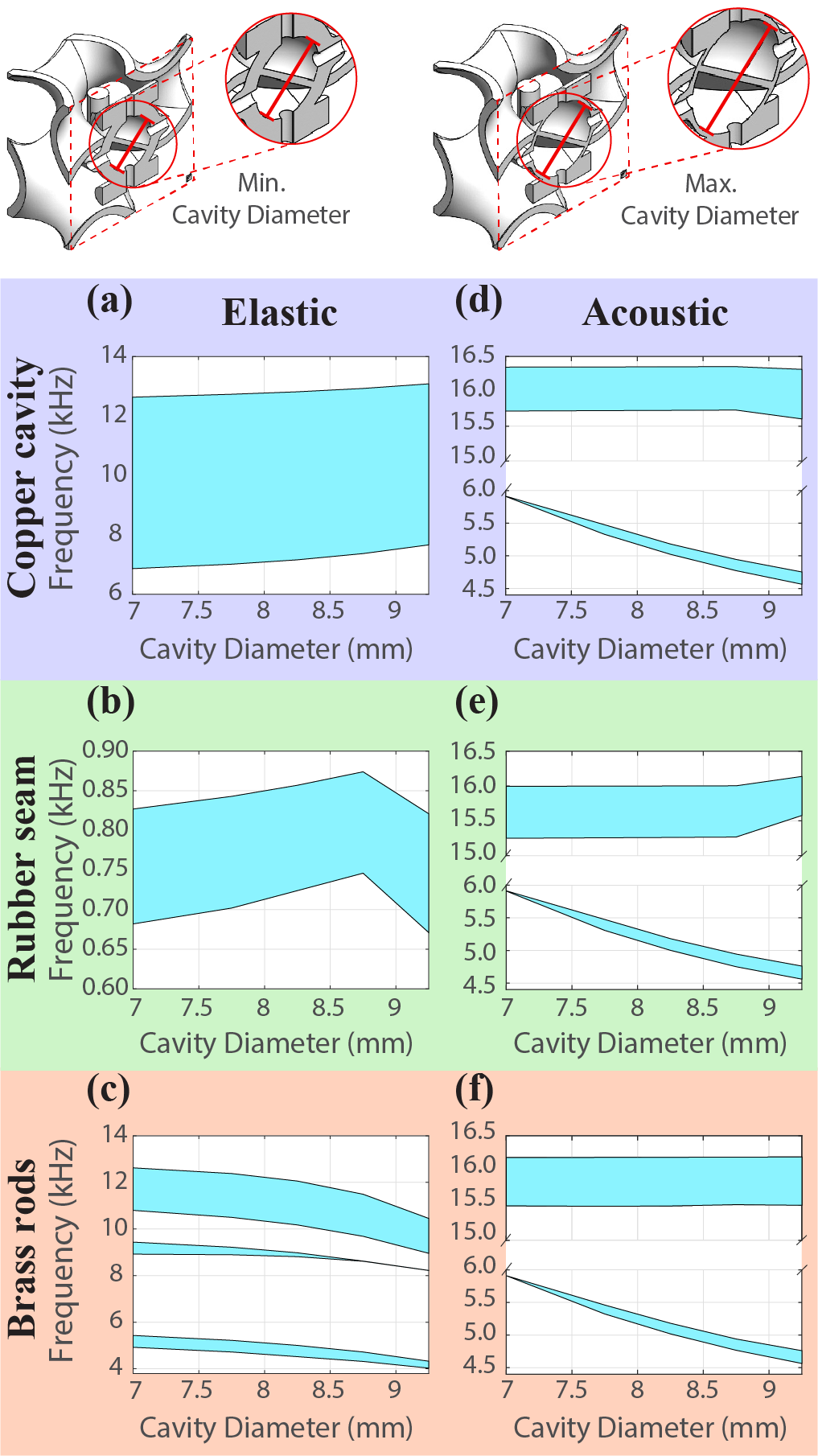}%
\caption{\label{fig:Inner Cutout Diameter Comparisons}\textbf{Helmholtz interior cavity diameter parametric comparisons.} (a) Copper cavity design band gap region in the elastic domain. (b) Rubber seam design band gap region in the elastic domain. (c) Brass rods design band gap region in the elastic domain. (d) Copper cavity design band gap region in the acoustic domain. (e) Rubber seam design band gap region in the acoustic domain. (f) Brass rods design band gap region in the acoustic domain.}
\end{figure}

By changing the thickness of the gyroid structure, we can tune the elastic wave attenuation of our designs, without altering its acoustic attenuation properties. The opposite can be achieved by changing the diameter of the opening to the air chamber (i.e., tuning the acoustic performance while preserving the same vibration attenuation behavior). Next, we explore tuning both fields (i.e., elastic and acoustic), simultaneously, by varying the diameter of the air chamber (Fig.\ref{fig:Inner Cutout Diameter Comparisons}). We vary the air chamber diameter from 7 to 9.25 mm. For elastic waves, the design with the copper cavity shows an increase of $\sim$ 0.5 kHz in the band gap central frequency, with the lower edge of the band gap increasing from $\sim$ 7 to 7.5 kHz (Fig.\ref{fig:Inner Cutout Diameter Comparisons}a). The upper edge of the band gap increases from $\sim$ 12.5 to 13 kHz. For the design with the rubber seam, the band gap central frequency increases by $\sim$ 0.5 kHz until the cavity diameter reaches 8.75 mm (Fig.\ref{fig:Inner Cutout Diameter Comparisons}b). Increasing the cavity diameter to 9.25 mm lowers the central frequency of the band gap region. Both the copper cavity and rubber seam elastic band gap frequency regions increase as the cavity diameter increases due to the decrease in mass of the central structure.  For the design with the brass rods, we observe the emergence of a new band gap between the previously seen gaps (Fig.\ref{fig:Inner Cutout Diameter Comparisons}c). As the cavity diameter increases, the band gap width decreases with the band gap central frequency decreasing by $\sim$ 1.5 kHz for the highest band gap and by $\sim$ 0.5 kHz in the lowest band gap. The middle band gap closes once the cavity diameter is greater than 8.75 mm. 

For acoustic waves, since all three designs are composed of the same cavity at the center of the unit cell, they have identical lower frequency resonance-based band gaps (Fig.\ref{fig:Inner Cutout Diameter Comparisons}d,e,f). As described by the Helmholtz resonator equation, volume and frequency are inversely proportional. Therefore, for all three designs, the band gap frequency region decreases as the cavity diameter increases, with the band gap central frequency decreasing by $\sim$ 1.25 kHz. When the cavity diameter is less than 7 mm, the band gap closes. The higher band gap frequency range is unaffected by the change in the cavity diameter in all three designs as the higher gap stems from scattering. The design with the rubber seam has a slight change in the higher band gap at 9.25 mm, increasing by $\sim$ 0.5 kHz on the lower edge of the band gap and by $\sim$ 0.25 kHz on the upper edge of the band gap.

The explored parameter space in this section shows the potential of tunability in either the elastic domain, the acoustic domain or both with the widest band gap spanning 60\%, the lowest elastic band gap starting at 367 Hz, and the lowest acoustic band gap starting at 1 kHz. The band gap percentage is defined as the band gap width divided by its central frequency \cite{bilal2012topologically}. However, the design space for our metamaterials is rich and the parameters of each design can be further optimized for specific operational frequencies and potential target applications.

\subsection{Experimental Results}

\begin{figure*}
\includegraphics{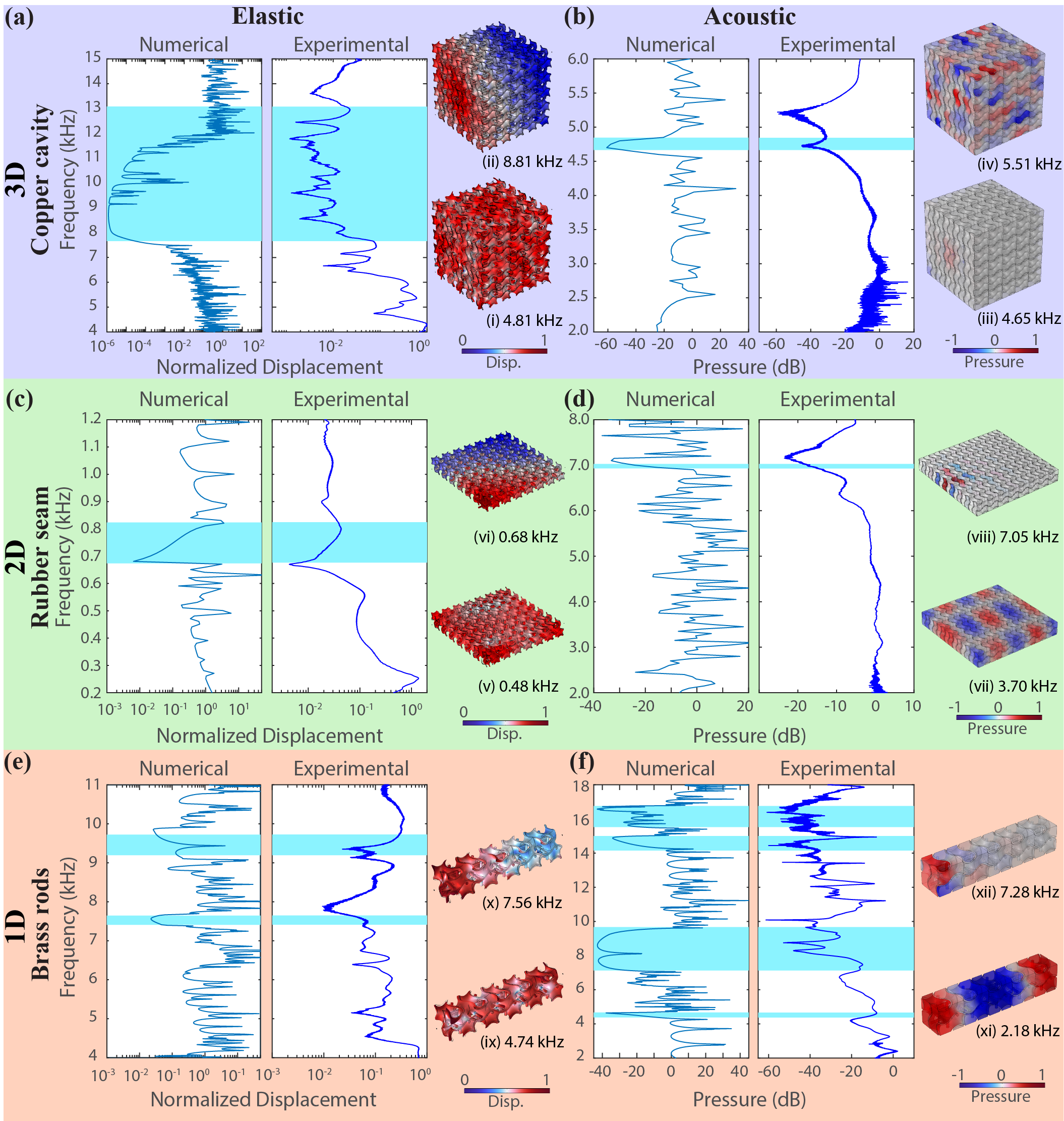}%
\caption{\label{fig:FRF Plots}\textbf{Numerical and experimental FRF comparisons.} (a) Copper cavity numerical versus experimental elastic FRFs with mode shapes. (b) Copper cavity numerical versus experimental acoustic FRFs with mode shapes. (c) Rubber seam numerical versus experimental elastic FRFs with mode shapes. (d) Rubber seam numerical versus experimental acoustic FRFs with mode shapes. (e) Brass rods numerical versus experimental elastic FRFs with mode shapes. (f) Brass rods numerical versus experimental acoustic FRFs with mode shapes.}
\end{figure*}

In order to validate our infinite unit cell numerical model, we consider a finite array of our metamaterials for all three proposed designs. We calculate the frequency response functions (FRFs) for our structures to confirm the band gap regions from the dispersion curves (which assumes infinite periodicity) but for finite metamaterial arrays.  In addition, we experimentally test our finite metamaterial arrays both in the elastic and acoustic domain.  For the elastic wave propagation experiments, we use a mechanical shaker attached to one end of the structures and measure the transmitted waves at the other end, using a scanning laser Doppler vibrometer (Polytech PSV 500). For the acoustic wave propagation experiments, we enclose the metamaterial arrays in a custom-built impedance tube and excite at one end by a loud speaker. We place microphones both before and after our samples to measure the acoustic wave transmission through our metamaterials. We excite the samples in both experiments using a chirp signal spanning the target operational frequency. To demonstrate the versatility of our design approach, we test our metamaterials, both numerically and experimentally with different dimensional periodicity. The copper-cavity design is tested with 3D periodicity (i.e., repeated in all directions). The rubber-seam design is tested with only two-dimensional (2D) periodicity (i.e., repeated only in the \textit{x-y} direction). The brass rods design is tested with only one-dimensional (1D) periodicity (i.e., repeated only in the \textit{x} direction)(Fig.\ref{fig:FRF Plots}). It is worth noting that we repeat our dispersion calculations in the case of 1D and 2D periodicity to account for the free edge(s) of the unit cell. These testing conditions show the utility of our metamaterial designs in scenarios where periodicity is only possible in certain directions or when volume constraints are present.  

We start the validation of our finite structural analysis for the copper-cavity design with 3D periodicity, we see confirmation of both elastic and acoustic band gap regions in the finite structure FRFs (Fig.\ref{fig:FRF Plots}a) as predicted by our dispersion analysis (Fig.\ref{fig:Dispersion Curves}). For elastic waves, the predicted wide band gap between 7.66 and 13.08 kHz is present in both our numerical simulations and experimental measurements reducing transmitted vibrations by $\sim$ two and half orders of magnitude. We note a slight reduction in the band gap width between the dispersion analysis and the numerical FRF. To help visualize the finite structure results, we plot the mode shapes at two different frequencies: (1) 4.81 kHz showing transmission throughout the structure, and (2) 8.81 kHz showing exponential decay of the displacement amplitude from excitation at the left edge to the right side of the sample. 

For acoustic waves, we observe the predicted band gap frequency range in both numerical calculations and experimental measurements (Fig.\ref{fig:FRF Plots}b). The realized sample shows noise reduction by $\sim$ 40 dB. The experimental result shows a second attenuation region centered around 5.25 kHz. We speculate that the additional dip in transmission stems from disorder, as the Helmholtz resonator cavities are assembled in the measured sample without perfect alignment of the hole openings of their respective cavities. In addition, we plot two mode shapes at a band gap frequency of 4.65 kHz and a pass band frequency at 5.51 kHz to better visualize the results. 

In the case of the rubber-seam design, we consider the repetition of the unit cells in the \textit{x-y} directions, forming a plate like structure. We excite the meta-plate at its corner for elastic wave simulation and measure at the opposing corner diagonally. We observe a clear band gap region around 0.7 kHz in both our numerical simulations and experimental measurements, aligning with our predicted band gap region from the dispersion curve with 2D periodicity. We show two mode shapes at a pass band frequency of 480 Hz and a band gap frequency at 680 Hz to better visualize the results (Fig.\ref{fig:FRF Plots}c). 

For acoustic wave simulations, we excite the center of the air portion of the design at the left edge, while assuming the solid part is rigid, and measure the transmission of the acoustic waves at the center of the opposing end. We observe clear attenuation at the predicted band gap frequency, centered around 7 kHz, in both our numerical and experimental FRFs. The realized sample shows noise reduction by $\sim$ 20 dB. We note that the realized sample has additional openings to the Helmholtz resonator cavity. These additional openings increase the band gap central frequency range  and are accounted for in the numerical models to better match the experimental results. In addition, we present two mode shapes to visualize both propagation at 3.70 kHz and attenuation at 7.05 kHz (Fig.\ref{fig:FRF Plots}d). 

In the case of the brass rod design, we consider the repetition of the unit cells only in the \textit{x} direction, forming a beam-like structure. In the elastic domain, we see general agreement between both numerical calculations and experimental measurements of the finite samples compared to our infinite unit cell analysis. The experimental measurement shows a slight up shift in the lower band gap frequency, likely due to the variation of the length of the brass rods in the manufactured sample. The upper band gap region shows good alignment with the predicted band gap frequency range. The mode shapes next to the FRFs show wave propagation at 4.74 kHz at a pass band frequency, and wave attenuation at 7.56 kHz inside the band gap frequency (Fig.\ref{fig:FRF Plots}e). 

For the acoustic waves, with the exception of the lowest frequency band gap, which shows a slight down shift in measurements, all other gaps are aligned between dispersion curves (with 1D periodicity), numerical FRF and experimental measurements. The mode shapes next to the FRFs show wave propagation at 2.18 kHz at a pass band frequency, and wave attenuation at 7.28 kHz within the second band gap frequency range (Fig.\ref{fig:FRF Plots}f). 

\section{Conclusion}
In conclusion, we present a methodology to augment gyroid based metamaterials, which have excellent stiffness-to-weight ratio, with multifunctional capabilities, namely: (1) impact mitigation, (2) vibration suppression, and (3) noise cancellation. We propose three different designs, all realized using different fabrication technologies: one with a copper cavity, another with a rubber seam, and a third with brass rods. The copper cavity design has the largest elastic band gap size at 60.39\% between 6.86 to 12.63 kHz. The rubber seam design has the lowest elastic band gap region at 41.61\% between 0.367 to 0.559 kHz. This design is the lightest of the three and has the best impact mitigation properties. The brass rods design has the highest stiffness-to-weight ratio of the three, and introduces two elastic band gap regions. All three designs are made with the same Helmholtz resonator parameters resulting in the same resonance acoustic band gaps, with the  lowest at 1 kHz. Further optimization of the proposed designs could lead to an even better performance depending on the application and the target performance metrics. Our findings can help incorporate metamaterials in practical applications that demand multi-functionality.

\section{Acknowledgment}
{This work was supported under Cooperative Agreement W56HZV-21-2-0001 with the US Army DEVCOM Ground Vehicle Systems Center (GVSC), through the Virtual Prototyping of Autonomy Enabled Ground Systems (VIPR-GS) program. DISTRIBUTION STATEMENT A. Approved for public release; distribution is unlimited. OPSEC9408}

\nocite{*}


%

\end{document}